\documentstyle[12pt]{article}
\begin{document}
E-print: hep-th/y0011188
\vskip1cm
\centerline{\large Rotating Super Black Hole as  Spinning Particle
\footnote{Talk given at the NATO Advanced Research Workshop
"Non-commutative Structures in Mathematics and Physics"
( Kiev, September 2000)}}
\vskip1cm
\par
\centerline{A. Ya. Burinskii}
\smallskip
\centerline{NSI of Russian Academy of Sciences }
\centerline{B. Tulskaya 52  Moscow 113191 Russia,
 e-mail:bur@ibrae.ac.ru}
\vskip1cm
\begin{quotation}
We give a review of the works devoted to the treatment of the Kerr super
black hole solution as a spinning particle.
The real, complex and stringy structures of the
Kerr and super-Kerr geometries are discussed, as well as the
recent results on the regular
matter source for the Kerr spinning particle.
\par
It is shown that the source has to represent a rotating
bag-like bubble having (A)dS interior and a smooth domain wall boundary.
    The given by Morris supersymmetric generalization
of the U(I) x U'(I) field model ( which was used by Witten to describe
cosmic superconducting strings ) is considered, and it is shown that this
model can be adapted for description of superconducting bags having a long
range external electromagnetic field and another gauge field confined
inside the bag.
\end{quotation}
PACS: 04.70, 12.60.Jr, 12.39.Ba, 04.65 +e
\newpage
\par
\section{Introduction}
\par
The Kerr rotating black hole solution
displays some remarkable features  indicating  a relation
to the structure  of   the   spinning   elementary particles.
In particular, in the 1969 Carter  \cite{car}  observed,
that  if  three parameters of the Kerr - Newman
metric are adopted to be ($\hbar $=c=1 ) $\quad e^{2}\approx  1/137,\quad
m \approx 10^{-22},\quad a \approx  10^{22},\quad ma=1/2,$
then one obtains  a  model  for  the four  parameters  of  the electron:
charge, mass, spin and  magnetic moment, and the gyromagnetic ratio is
automatically the same as that of the Dirac electron.
Investigations along this line \cite{is,bur,ibur,lop,burbag} allowed
to find out stringy structures in the real and complex Kerr geometry
and to put forward a conjecture on
the baglike structure of the source of the Kerr-Newman solution.
The earlier investigations  \cite{is,ham,lop} showed that this source
represents a rigid rotator ( a relativistic disk ) built of an exotic matter
with superconducting properties.
Since 1992 black holes have paid attention of string theory.
In 1992 the Kerr solution was generalized by
Sen to low energy string theory
\cite{sen}, and it was shown \cite{bursen} that near the Kerr singular ring
the Kerr-Sen solution acquires a metric similar to the field around a
heterotic string.
The point of view has appeared that  black holes can be treated as
elementary particles \cite{part}.
On the other hand, a description of a spinning particle based only on the
bosonic fields cannot be complete, and involving fermionic degrees of
freedom is required. Therefore, the  spinning particle must be based
on a super-Kerr-Newman black hole solution \cite{superBH} representing a
natural combination of the Kerr spinning particle and superparticle model.
Angular momentum $L$ of spinning particles is very high
$\mid a\mid  = L/m \geq  m$, and the horizons of the Kerr metric disappear.
There appears a naked  ring-like  singularity which has to be regularized
being replaced by a smooth matter source.
In this review we consider a source representing a rotating superconducting
bag with a smooth domain wall boundary described by
a supersymmetric version of the $U(I)\times U^\prime (I)$ field model
\cite{Mor}.
In fact, this model of the Kerr-Newman source
represents a generalization of the Witten superconducting
string model \cite{Wit} for the superconducting baglike sources \cite{burbag}.
\section{ Complex source of Kerr geometry and its stringy interpretation}
The Kerr-Newman solution can be represented in the Kerr-Schild form
 \begin{equation}
g_{\mu\nu} =
\eta_{\mu\nu} + 2 h e^3_{\mu} e^3_{\nu}, \label{KSf} \end{equation}
where $\eta _{\mu\nu}$ is metric of an auxiliary Minkowski space
 $\eta_{\mu\nu}= diag (-1,1,1,1)$, and $h$ is a scalar
function. Vector field $ e^3 $ is null,
$ e^3_{\mu} e^{3 \mu}=0, $ and tangent to PNC ( principal null congruence )
of the Kerr geometry. The Kerr PNC is twisting
i.e. corresponding to a vortex of a null radiation.
\footnote{Besides, the Kerr PNC is
geodesic and shear free, it represents a bundle of twistors and can be
described by the Kerr theorem \cite{dks,bkp,c-str,pla}.}
One of the main peculiarities of the Kerr geometry is singular ring
representing a branch line of the Kerr space on the `positive' ($ r>0 $) and
`negative'($ r<0$ ) sheets which are divided by the disk $r=0$ spanned by
this ring. The Kerr singular ring is exhibited as a pole of the function
$h(r,\theta) = \frac{mr -e^2/2}{r^2 + a^2 \cos ^2 \theta},$
where $r$ and $\theta$ are the oblate spheroidal coordinates.
The Kerr PNC is in-going on the `negative' sheet of space, it crosses
the disk $r=0$ and turns into out-going one on the `positive' sheet.
The simplest solution possessing the Kerr singular ring
was obtained by Appel in 1887 (!) \cite{app}. It can be considered
as a Newton or a Coulomb analogue to the Kerr solution.
On the real space-time the singular ring arises in the Coulomb solution
$f=e/\tilde r$,  where
$ \tilde {r} = \sqrt{ (x-x_o)^2+(y-y_o)^2 +(z-z_o)^2} $,
when the point-like source is shifted to a complex point of space
$(x_o, y_o , z_o ) \rightarrow (0,0,ia)$. Radial distance
$\tilde r $ is complex in this case and can be expressed
in the oblate spheroidal coordinates
$r$ and $\theta$ as $\tilde r = r + i a \cos \theta$.
The source of Kerr-Newman solution, like the Appel solution,
can be considered from complex point of view as a "particle" propagating
along a complex world-line \cite{c-str,pla} parametrized by complex time.
\par
  The objects described by the
complex world-lines occupy an intermediate position between  particle
and  string. Like the string
they form the two-dimensional surfaces or the world-sheets in the
space-time. It was shown that the complex Kerr source may be considered as
a complex hyperbolic string which requires an orbifold-like structure of
the world-sheet. In many respects this source is similar to the `mysterious'
$N=2$ string of superstring theory shedding a light on the puzzle of its
physical interpretation
As we have already mentioned, there is one more stringy structure in the
Kerr geometry connected with the Kerr singular ring. In fact the both these
stringy structures are different exhibitions of some membrane-like source.
This source has a complex interpretation alongside with a real image in
the form of a rotating bubble which will be discussed further.
\par
The  Kerr PNC may be obtained from the complex
source by a retarded-time construction.
The rays of PNC are the tracks of null planes of the complex light cones
emanated from the complex world line \cite{bkp,pla}.
The complex light cone with the vertex at some point
$x_0$ of the
complex world line $x_0^\mu(\tau)$:
$(x_\mu - x_{0 \mu})(x^\mu -x_0^\mu) = 0 ,$
can be split into two families of  null planes:  "left" planes
$ x_L = x_0(\tau) + \alpha e^1 + \beta e^3 $
spanned by null vectors $e^1$ and $e^3$,
and"right"planes
$ x_R = x_0(\tau) + \alpha e^2 + \beta e^3,$
spanned by null vectors $e^2$ and $e^3$.
\footnote{The Kerr's tetrad null vector-forms are: $e^1 = d \zeta - Y d v,
\quad e^2 = d \bar \zeta - \bar Y d v, \quad
e^3 = du + \bar Y d \zeta + Y d \bar \zeta - Y\bar Y dv, \quad
e^4 = dv + h e^3$,
where the Cartesian null coordinates $u,v,\zeta,\bar \zeta$ are used.}
 The Kerr PNC arises as the real slice of the family of the "left" null
planes of the complex light cones which vertices lie on the straight complex
world line $x_0(\tau)$.
\par
Only the cones lying on the strip $\vert Im \tau \vert \le \vert a \vert$
have a real slice. Therefore, the ends of the resulting complex string
are open. To satisfy the complex boundary conditions, an orbifold-like
structure of the worldsheet must be introduced \cite{c-str,pla},
which is closely connected with the above mentioned Kerr's twosheetedness.
\section{Super-Kerr-Newman geometry}
A supergeneralization of the Kerr-Newman solution can be
obtained as a natural combination of the Kerr spinning particle and
superparticle \cite{superBH}.
In fact, the complex structure of the Kerr geometry
suggests the way of its supergeneralization.
\par
Note, that any exact solution of the Einstein gravity
is indeed a trivial solution of supergravity field equations.
The supergauge freedom allows one to turn any gravity solution into a form
containing spin-3/2
field $\psi_\mu$ satisfying the supergravity field equations. However, since
this spin-3/2 field can be gauged away by the reverse transformation,
such supersolutions have to be considered as {\it trivial}.
The hint how to avoid this triviality problem follows from the complex
structure of the Kerr geometry. In fact, from the complex point of view
the Schwarzschild and Kerr geometries
are equivalent and connected by a {\it trivial} complex shift.
\par
The {\it non-trivial} twisting structure of the Kerr geometry arises as a
result of the complex {\it shift of the real slice} concerning the center
of the solution \cite{bkp,c-str}.
Similarly, it is possible to turn a {\it trivial} super black hole
solution into a {\it non-trivial}.
 The {\it trivial supershift} can be represented as a
replacement of the complex world line by a superworldline
$X^\mu _0(\tau)= x^\mu _0 (\tau)-i \theta \sigma ^\mu \bar \zeta +i\zeta
\sigma^\mu \bar \theta,$
parametrized by Grassmann coordinates
$\zeta, \quad \bar \zeta$, or as a corresponding coordinate
replacement in the Kerr solution
\begin{equation}
x^{\prime \mu}  = x^\mu + i \theta\sigma^\mu \bar \zeta
 - i \zeta\sigma^\mu \bar \theta;
\qquad
\theta^{\prime}=\theta + \zeta ,\quad
{\bar\theta}^{\prime}=\bar\theta + \bar\zeta, \label{SG}
\end{equation}
\par
Assuming that coordinates $x^\mu$ before the supershift were the usual
c-number coordinates one sees that coordinates acquire nilpotent
Grassmann contributions after supertranslations. Therefore, there
appears a natural splitting of the space-time coordinates on the
c-number `body'-part and a nilpotent part - the so called `soul'.
The `body' subspace of superspace, or B-slice, is a submanifold
where the nilpotent part is equal to zero,
and it is a natural analogue to the real slice of the complex case.
\par
Reproducing the real slice procedure of the Kerr geometry in superspace
one has to use the replacements:
\par
- complex world line $\to$ superworldline,
\par
- complex light cone $\to$ superlightcone,
\par
- real slice $\to$ body slice.
\par
Performing the body-slice procedure to superlightcone constraints
\begin{equation}
s^2= [x_\mu -X_{0\mu}(\tau)][x^\mu - X^\mu _0(\tau)] =0 ,
\label{slc}\end{equation}
one selects the body and nilpotent parts of this equation and obtains three
equations. The first one is the discussed above real slice condition of the
complex Kerr geometry claiming that complex light cones can reach the real
slice. The nilpotent part of (\ref{slc}) yields two B-slice conditions
\begin{equation} [x^\mu -x_0^\mu (\tau)]
( \theta\sigma_\mu \bar \zeta
 - \zeta\sigma_\mu \bar \theta)=0; \label{odd1}\end{equation}
\begin{equation}
( \theta\sigma \bar \zeta
 - \zeta\sigma \bar \theta)^2 =0.\label{odd2}\end{equation}
These equations can be resolved by representing the complex light cone
equation via the commuting two-component spinors $\Psi$ and $\tilde \Psi$:
$x_\mu = x_{0\mu} + \Psi \sigma _\mu \tilde \Psi .$
"Right" (or "left")  null planes of the complex light cone can be obtained
keeping  $\Psi$ constant and varying  $\tilde \Psi$ (or keeping
$\tilde \Psi$ constant and varying  $ \Psi$.)
As a result we obtain the equations
$\bar\Psi \bar\theta =0,\qquad\bar\Psi \bar\zeta =0,$
which in turn are  conditions of proportionality of the commuting spinors
$\bar\Psi(x)$  determining the PNC of the Kerr geometry
and anticommuting spinors $ \bar\theta$ and $\bar\zeta$,
these conditions providing the left  null superplanes of the supercones to
reach B-slice.
It also leads to $ \bar\theta \bar\theta= \bar\zeta \bar\zeta=0, $
and equation (\ref{odd2}) is satisfied automatically.
\par
Thus, as a consequence of the B-slice and superlightcone constraints
we obtain a non-linear submanifold of superspace $\theta = \theta (x),
\quad \bar \theta = \bar \theta (x).$
The original four-dimensional
supersymmetry is broken, and the initial supergauge freedom  which
allowed to turn the super geometry into trivial one is lost. Nevertheless,
there is a residual supersymmetry based on free Grassmann parameters
$ \theta ^1, \quad \bar \theta ^1$.
\par
The above B-slice constraints
yield in fact the non-linear realization of broken supersymmetry introduced
by Volkov and Akulov \cite{VA,WB} and considered in N=1 supergravity by
Deser and Zumino \cite{DZ}.
It is assumed that this construction is similar to the Higgs
mechanism of the usual gauge theories and
$\zeta ^\alpha (x), \quad \bar \zeta ^{\dot\alpha} (x) $
represent  Goldstone fermion which  can be eaten by appropriate local
supertransformation
$\epsilon (x)$ with  a corresponding redefinition of the tetrad
and spin-3/2 field. Complex character of supertranslations in the Kerr case
demands to use in this scheme the N=2 supergravity \cite{FN}.
We omit here details referring to \cite{superBH} and mention
only that
in the resulting exact solution the torsion and Grassmann contributions to
tetrad cancel, and metric takes the exact Kerr-Newman form.
However there are the extra wave fermionic fields on the bosonic Kerr-Newman
background propagating along the Kerr PNC and concentrating near the Kerr
singularity (traveling waves). Solution contains also an
extra axial singularity which is coupled topologically with singular
ring threading it.
\section{Baglike source of the Kerr-Newman solution}
 The above consideration of super-Kerr-Newman solution is based on the
massless fields providing description of the rotating super-black-hole.
It could be the end of story since the source of a rotating black hole is
hidden behind the horizons.
\par
However, the value of angular momentum for spinning particles
is very high regarding the mass parameter and the horizons disappear
uncovering the Kerr singular ring. To get a regularized solution the
massless fields of the black hole solution have to get a mass in the core
region forming a matter source removing the Kerr singularity and
twosheetedness of the Kerr space.\footnote{This problem is
actual for black hole physics, too. See for example \cite {DEH} and
references therein.}
\par
Obtaining a regularized Kerr source represents an old problem.
In the first disk-like model given by Israel \cite{is} a
truncation of the negative sheet was used.
As a result there appeared a source distribution on the
surface of the disk $r=0$. Analyzing the resulting stress-energy tensor
Hamity showed \cite{ham} that this disk has to be in a rigid relativistic
rotation and built of an exotic matter having zero energy density and
negative pressure.
In the development of this model given by L\'opez \cite{lop} the truncation
is placed at the coordinate surface $r=r_e=\frac{e^2}{2m}$ ( where $h=0$ ),
and the region $r<r_e$ is replaced by Minkowski space. As a result the
 source takes the form of the highly oblate and infinitely thin elliptic
shell of the Compton radius $a =\frac{1}{2m}$ and of the thickness of the
classical Dirac electron radius $r_e$.
For small angular momentum the source takes the form of the
Dirac electron model, a charged sphere of the classical size $r_e$.
The fields out of the shell have the exact Kerr-Newman form. Interior of the
shell is flat. The shell is charged and rotating, and built of a
superconducting matter. In corotating space one sees that matter has a
negative pressure and zero energy density.
\par
The L\'opez source represents a bubble with an infinitely thin domain wall
boundary.
In the paper \cite{burbag} an attempt was undertaken to get the source of
the Kerr-Newman solution with a smooth matter distribution.
\footnote{First attempts in this direction were undertaken in the
papers \cite{lop1,GG}.}
Retaining the metric in the Kerr-Schild form (\ref{KSf}) and the form ( and
main properties ) of the Kerr PNC, it was assumed that function
$h(r,\theta)$
takes a more general form $ h =\frac {f(r)}{r^2+a^2 \cos ^2 \theta}$,
where the function $f(r)$ is continuous and takes the usual
Kerr-Newman form $f_{KN}(r)=mr-e^2/2$ in the external region. In the same
time, in a neighborhood of the Kerr disk $r\le r_0$ ( the core region )
including the Kerr singularity, the function $f(r)$ has to satisfy
some conditions of regularity to provide finiteness of
the metric and the stress-energy tensor of source, which is determined by
the Einstein equations for this metric.
\par
It was shown that this regularity is achieved for
the function $f(r)\sim r^n$ with $n\ge 4$.
In the case $n=4$, $f(r)=f_0(r)=\alpha r^4$,
( in the nonrotational case $ a=0 $ ) space-time has a constant curvature
in the core and generated by a homogenous matter distribution with energy
density $\rho= \frac 1{8\pi} 6\alpha$.  Therefore, assuming that matter in
the core has a homogenous distribution one can estimate the
boundary of the core region $r_0$
as a point of intersection of $f_0(r)$ and  $f_{KN}(r)$.
Regularity of the stress-tensor demands continuity of the function $f(r)$
up to first derivative, therefore, the resulting smooth function $f(r)$ must
be interpolating between functions $f_0(r)$ and $f_{KN}(r)$ near the
boundary of the core $r\approx r_0$.
\par
Let us now mention that general metric
(\ref{KSf}) can be expressed via orthonormal tetrad as follows \cite{burbag}
\begin{equation}
g_{\mu\nu} =m_\mu m_\nu + n_\mu n_\nu + l _\mu l_\nu - u_\mu u_\nu,
\label{gBL}
\end{equation}
and the corresponding stress-energy tensor of the source ( following from the
Einstein equations ) may be represented in the form
\begin{equation}
T_{\mu\nu}^{(af)} = (8\pi)^{-1} [(D+2G) g_{\mu\nu} -
(D+4G) (l_\mu l_\nu -  u_\mu u_\nu)],
\label{Taf}
\end{equation}
where $u_\mu$ is the unit time-like four-vector,
$l_\mu $ is the unit vector in radial direction, and
$n_\mu, m_\mu $
are two more space-like vectors.
Here
\begin{equation}
 D= - f^{\prime\prime}/(r^2 + a^2 \cos ^2 \theta),
\label{D}
\end{equation}
\begin{equation}
G= (f'r-f)/ (r^2 + a^2 \cos ^2 \theta)^2,
\label{G}
\end{equation}
and
the Boyer-Lindquist coordinates $t,r,\theta, \phi$ are used.
The expressions (\ref{Taf}), (\ref{D}), (\ref{G})
show that the source represents a smooth
distribution of rotating confocal ellipsoidal layers $r=const.$.
\par
Like to the results for singular (infinitely thin) shell-like source
\cite{ham,lop},
the stress-energy tensor can be diagonalized in a comoving
coordinate system showing that the source represents a relativistic rotating
disk. However, in this case, the disk is separated into
ellipsoidal layers each of which rotates rigidly with its own angular
velocity $\omega (r) =a/(a^2+r^2)$.
In the comoving coordinate system the tensor $T_{\mu\nu}$ takes the form
\begin{equation}
T_{\mu\nu} =  \frac{1}{8\pi}
\left( \begin{array}{cccc}
2G&0&0&0 \\
0&-2G&0&0 \\
0&0&2G+D&0 \\
0&0&0&2G+D
\end{array} \right),
\label{Tful}
\end{equation}
that corresponds to energy density
$\rho = \frac{1}{8\pi} 2G$,
 radial pressure $p_{rad} = -\frac{1}{8\pi} 2G$,
and tangential pressure $p_{tan} = \frac{1}{8\pi}(D+ 2G)$.
\footnote{Similar interpretation was also given in \cite{lop1}, however,
the expression for stress-energy does not
contain very important $D$ term there.}
\par
Setting $a=0$ for the non-rotating case, we obtain
$\Sigma= r^2$, the surfaces $r=const.$ are spheres and we have spherical
symmetry for all the above relations. The region described by
$f(r)=f_0(r)$ is the region of constant value of the scalar curvature
invariant $R=2D= - 2 f_0^{\prime\prime}/r^2 = - 24\alpha$, and of a constant
value of energy density.
If we assume that the region of a constant curvature is closely extended
to the boundary of source $r_0$ which is determined as a root of
the equation
\begin{equation}
f_0(r_0)=f_{KN}(r_0),
\label{r0}
\end{equation}
then, smoothness of the $f(r)$ in a
small neighborhood of $r_0$, say $\vert r-r_0 \vert <\delta$, implies
a smooth interpolation  for the derivative of the function $f(r)$
between $f^\prime _0(r)\vert _{r=r_0-\delta}$ and
$f^\prime _{KN}(r)\vert_{r=r_0+\delta}$.
Such a smooth interpolation on a small distance $\delta$ shall
lead to a shock-like increase of the second derivative
$f^{\prime\prime} (r)$ by $ r \approx r_0 $.
\par
In charged case for $\alpha \le 0$ ( AdS internal geometry of core) there
exists only one positive root $r_0$,
and second derivative of the smooth function $f^{\prime\prime} (r)$
is positive near this point. Therefore, there appears an extra tangential
stress near $r_0$ caused by the term
$D=- f^{\prime\prime} (r)/(r^2 + a^2 \cos ^2 \theta )\vert _{r=r_0}$
in the expression (\ref{Tful}).
It can be interpreted as the appearance of an effective shell ( or a domain
wall ) confining the charged ball-like source with a geometry of a constant
curvature inside the ball.
The case
$\alpha=0$ represents the bubble with a flat interior which has
in the limit $\delta\to 0$ an infinitely thin shell. It
corresponds to the L\'opez model.
\par
The point $r_e=\frac {e^2}{2m}$ corresponding to a ``classical size" of
electron is a peculiar point as a root of the equation $f_{KN}(r) =0$.
 It should be noted that by $\alpha=0$ the equation (\ref{r0}) yields
the root $r_0=r_e$. For $\alpha <0$ position of the roots is $r_0<r_e$,
and for $\alpha >0$ one obtains $r_0> r_e $.
Therefore, there appear four parameters characterizing the function
$f(r)$ and the corresponding bag-like core: parameter $\alpha$ characterizing
cosmological constant inside the bag $\Lambda _{in}= 6 \alpha$,
two peculiar points $r_0$ and $r_e$ characterizing the size of the bag and
parameter $\delta$ characterizing the smoothness of the function $f(r)$ or
the thickness of the domain wall at the boundary of core.
\par
The internal geometry of the ball is de Sitter one for $\alpha>0$,
anti de Sitter one for $\alpha <0$ and flat one for $\alpha=0$.
\par
Let us consider peculiarities of the rotating Kerr source.
In this case the
surfaces $r=const.$ are ellipsoids described by the equation
$\frac{x^2 +y^2}{r^2 + a^2} + \frac {z^2}{r^2} =1.$
Energy density inside the core will be constant only in the equatorial plane
$\cos \theta=0$. Therefore, the Kerr singularity is regularized and the
curvature is constant in string-like region $r<r_0$ and $\theta=\pi/2$
near the former Kerr singular ring.
The ratio
$\frac{stress\vert _{\theta=0}}{stress \vert_{\theta=\pi/2}}
< (r_e/ a)^4 = e^8 < 10^{-8}$ shows a
strong increase of the stress near the string-like boundary of the disk.
\subsection{Field model for the bag: From superconducting strings to
superconducting bags}
The known models of the bags and cosmic bubbles
 with smooth domain wall boundaries are based on the Higgs
scalar field $\phi$  with a Lagrange density of the form
$L=- \frac 12 \partial _\mu \phi\partial ^\mu \phi
-\frac {\lambda ^2}8 (\phi ^2-\eta^2)^2 $ leading to the kink
planar solution ( the wall is placed in $xy$-plane at $z=0$ )
\begin{equation}
\phi (z)= \eta \tanh (z/\delta),
\label{kink}
\end{equation}
where $\delta =\frac 2{\lambda \eta}$ is the wall thickness.
The kink solution describes two topologically distinct vacua
$<\phi>= \pm \eta$ separated by the domain wall.
\par
The stress--energy
tensor of the domain wall is
\begin{equation}
T_{\mu}^{\nu} = \frac{\lambda ^2 \eta ^4}4 \cosh ^{-4}(z/\delta)
diag (1,1,1,0),
\label{Tkink}
\end{equation}
indicating a surface stress within the plane of the wall which is equal
to the energy density.
When applied to the spherical bags or cosmic bubbles \cite{Mac,MB},
the thin wall
approximation is usually assumed $\delta \ll r_0$, and a spherical domain
wall separates a false vacuum inside the ball ($r<r_0$) $<\phi>_{in}=-\eta$
from a true outer vacuum $<\phi>_{out}=\eta$.
\par
In the gauge string models, the Abelian Higgs field provides
confinement of the magnetic vortex lines in superconductor. Similarly,
in the models of superconducting bags,
the gauge Yang-Mills or quark fields are confined
in a bubble ( or cavity ) in superconducting QCD-vacuum.
\par
A direct application of the Higgs
model for modelling superconducting properties of the Kerr source
is impossible since the Kerr source has to contain the external long range
Kerr-Newman electromagnetic field, while in the models of strings and bags
the situation is quite opposite: vacuum is superconducting in external
region and electromagnetic field acquires a mass there from Higgs field
turning into a short range field.
An exclusion represents the  $U(I)\times \tilde U(I)$ cosmic string model
given by Vilenkin-Shellard and Witten \cite{VS,Wit}  which
represents a doubling of the usual Abelian  Higgs model.
The model  contains
two sectors, say $A$ and $B$, with two Higgs fields $\phi_A$ and $\phi_B$,
and two gauge fields $A_{\mu}$ and $B_{\mu}$ yielding two sorts
of superconductivity $A$ and $B$.
It can be adapted to the bag-like source in such a manner
that the gauge field $A_{\mu}$ of the $A$ sector has to describe a
long-range electromagnetic field in outer region of the bag
while the chiral scalar field of this sector $\phi_A$ has to form a
superconducting core inside the bag which must be unpenetrable
for $A_{\mu}$ field.
\par
The sector $B$ of the model has to describe the opposite situation.
The chiral field $\phi_B$ must lead to a $B$-superconductivity in outer
region confining the gauge field $B_{\mu}$ inside the bag.
\par
The corresponding Lagrangian of the Witten $U(I)\times \tilde U(I)$
field model is given by
\cite{Wit}
\begin{equation}
L=-(D^\mu \phi_A )(\overline{ D_\mu \phi_A} )-(\tilde D^\mu \phi_B )
(\overline{\tilde D_\mu \phi_B} )-\frac 14 F_A^{\mu \nu }F_{A\mu \nu }-
\frac 14F_B^{\mu \nu } F_{B\mu \nu}-V,
\label{L}
\end{equation}
where
$F_{A\mu \nu }=\partial _\mu A_\nu -\partial _\nu A_\mu $ and
$F_{B\mu \nu}=\partial _\mu B_\nu -\partial _\nu B_\mu $
are field stress tensors, and the potential has the form
\begin{equation}
V=\lambda (\bar{\phi_B}\phi_B -\eta ^2)^2+f(\bar{\phi_B}
\phi_B -\eta ^2)\bar{\phi_A}\phi_A +m^2\bar{\phi_A}\phi_A +\mu
(\bar{\phi_A}\phi_A )^2.  \label{V}
\end{equation}
Two Abelian gauge fields $A_\mu$ and $B_\mu$ interact separately
with two complex scalar fields $\phi _B$ and $\phi_A$ so that the
covariant derivative $D_\mu \phi_A =(\partial +ie A_\mu) \phi_A $ is
associated with $A$ sector, and covariant derivative
$\tilde D_\mu \phi _B = ( \partial +ig B_\mu) \phi _B $ is associated with
$B$ sector.
The model fully retains the properties of the usual bag models
which are described by $B$ sector providing confinement of $B_{\mu}$ gauge
field inside bag, and it acquires the long range electromagnetic field
$A_{\mu}$ in the outer-to-the-bag region described by sector $A$.
The A and B sectors are almost independent interacting only
through the potential term for scalar fields. This interaction has to
provide synchronized phase transitions from superconducting B-phase
inside the bag to superconducting A-phase in the outer region.
The synchronization of this transition occurs explicitly in a supersymmetric
version of this model given by Morris \cite{Mor}.
\subsection{Supersymmetric Morris model}
In Morris model, the main part of Lagrangian of the bosonic sector is
similar to the Witten field model. However,  model
has to contain an extra scalar field $Z$ providing  synchronization of
the phase transitions in $A$ and $B$ sectors.~\footnote{In fact the Morris
model contains fife complex chiral fields
$\phi _i=\{ Z, \phi _-,\phi _+, \sigma _-, \sigma _+ \} $. However,
the following identification of the fields is assumed $ \phi =\phi _+;\;
\bar \phi = \phi _-$ and $ \sigma = \sigma _+;\; \bar \sigma =\sigma_-$.
In previous notations $\phi \sim\phi_A$ and $\sigma \sim \phi _B$.}
\par
The effective Lagrangian of the Morris model has the form
\begin{eqnarray}
L=-2(D^\mu \phi )\overline {( D_\mu  \phi  )}-2(\tilde D^\mu \sigma )
(\overline {\tilde D_\mu \sigma} )-\partial ^\mu Z \partial _\mu
\bar Z \nonumber \\
-\frac 14F^{\mu \nu }F_{\mu \nu }-
\frac 14 F_B ^{\mu \nu }F_{B\mu \nu}-V (\sigma, \phi, Z),
\label{SL}
\end{eqnarray}
where the potential $V$ is determined through the superpotential $W$ as
\begin{equation}
V= \sum _{i=1} ^5 \vert W_i \vert ^2=2\vert\partial W /\partial \phi\vert^2 +
2\vert\partial W /\partial \sigma \vert^2 +
\vert\partial W /\partial Z \vert^2 .
\label{SV}
\end{equation}
The following superpotential, yielding the gauge invariance and
renormalizability of the model, was suggested
\footnote{Superpotential is holomorfic function of
$\{Z,\phi, \bar\phi, \sigma,
\bar \sigma\}$.}
\begin{equation}
W=\lambda {Z}(\sigma \bar\sigma -\eta ^2) + ( c Z+m ) \phi\bar \phi,
\label{SW}
\end{equation}
where the parameters $\lambda $, $c$, $m$, and $\eta $ are real
positive quantities.
\par
The resulting scalar potential $V$ is then given by
\begin{eqnarray}
V= & \lambda ^2(\bar \sigma \sigma -\eta ^2)^2
+2\lambda c(\bar\sigma\sigma -\eta ^2)\phi \bar\phi
+c^2(\bar\phi \phi)^2  + \\
&2 \lambda ^2 \bar Z Z \bar\sigma \sigma
+2(c\bar Z +m)(cZ+m)\bar\phi \phi.  \nonumber
\label{Vfull}
\end{eqnarray}
\subsubsection{Supersymmetric vacua}
>From (\ref{SV}) one sees that the supersymmetric vacuum states,
corresponding to the lowest value of the potential,
are determined by the conditions
\begin{eqnarray}
F_\sigma = - \partial \bar W / \partial \bar \sigma =0; \\
F_\phi = - \partial \bar W / \partial \bar \phi =0;\\
F_Z = - \partial \bar W / \partial \bar Z  =0,
\label{Svac}
\end{eqnarray}
and yield $V=0$.
These equations lead to two supersymmetric vacuum states:
\begin{equation}
I ) \qquad Z=0;\quad \phi=0 ;\quad \vert\sigma\vert=\eta ;\quad W=0;
\label{true}
\end{equation}
and
\begin{equation}
II ) \qquad Z=-m/c;\quad \sigma=0; \quad \vert \phi \vert =\eta
\sqrt{\lambda/c};\quad W=\lambda m\eta ^2/c.
\label{false}
\end{equation}
We shall take the state $I$ for external region of
the bag, and the state $II$ as a state inside the bag.
\par
The treatment of the gauge field $A_\mu$ and $B_\mu$ in $B$ is similar
in many respects because of the symmetry between  $A$ and $B$ sectors
allowing one to consider the state $\Sigma = \eta$ in outer region as
superconducting one in respect to the gauge field $B_\mu$.
Field $B_\mu$ acquires the mass $m_B= g \eta $ in outer region, and the
$\tilde U(I)$ gauge symmetry is broken, which provides confinement of the
$B_{\mu}$ field inside the bag.
The bag can also be filled by quantum excitations of fermionic,
or non Abelian fields.
The interior space of the Kerr bag is regularized in this model since the
Kerr singularity and twofoldedness are suppressed by function $f=f_0(r)$.
However, a strong increase of the fields near the former Kerr singularity
can be retained leading to the appearance of traveling waves along the
boundary of the disk.
\par
\subsection{Supersymmetric bubble based on the Morris field
model}
It is shown in \cite{burbag} that in the planar thin wall approximation,
and by neglecting the gauge fields there is a supersymmetric BPS-saturated
domain wall solution interpolating between supersymmetric vacua I) and II).
This domain wall displays the usual structure of stress-energy tensor with
a tangential stress.
The non-zero components of the stress-energy tensor take the form
\begin{eqnarray}
T_{00} & = &-T_{xx} =- T_{yy}=\frac{1}{2}[ \delta _{ij}( \Phi ^i,_z)
(\Phi ^j,_z) + V];\\
T_{zz}& = &\frac{1}{2}[ \delta_{ij}( \Phi ^i,_z)(\Phi ^j,_z) - V],
\label{Tflat}
\end{eqnarray}
where $\Phi _i=\{ Z, \phi _-,\phi _+, \sigma _-, \sigma _+ \} $.
One can estimate the  mass and energy of a bubble formed by such
a domain
wall in global supersymmetry setting vacuum I) as external one and
vacuum II) as an internal vacuum.
Using the Tolman relation
$M =\int dx^3 \sqrt{-g}(-T_0^0+T_1^1 +T_2^2 +T_3^3)$,
replacing coordinate $z$ on radial coordinate $r$, and integrating over
sphere one obtains
\begin{equation}
M_{bubble}= -4\pi \int V(r) r^2 dr = -4\pi \int (\Phi ^i,_r)^2 r^2 dr .
\label{Mwall}
\end{equation}
The resulting effective mass is negative, which is caused by gravitational
contribution of the tangential stress. The repulsive
gravitational field was obtained in many singular and
smooth models of domain walls \cite{IpsSik,lop2,CvSol,CG}.
One should note, that similar gravitational contribution to the mass caused
by interior of the bag will be $M_{gr.int}=\int D  r^2 dr
=-\frac 23 \Lambda r_0^3$. It depends on the sign of curvature inside
the bag and will be negative in de Sitter case and positive in AdS one.
\par
The total energy of a uncharged bubble
forming from the supersymmetric BPS saturated domain wall is
\begin{equation}
E_{0 bubble} = E_{wall} =
4\pi \int _0 ^\infty \rho r^2 dr \approx 4\pi r_0 ^2
\epsilon_{min},
\label{E0tot}
\end{equation}
where $r_0$ is radius of the bubble, and $$\epsilon_{min} =
W(0)-W(\infty) =\lambda m \eta^2/c.$$
Corresponding total mass following from the Tolman relation will be
negative
\begin{equation}
M_{0 bubble} = - E_{wall} \approx -4\pi r_0 ^2 \epsilon_{min}.
\label{M0tot}
\end{equation}
It is the known fact showing that the uncharged bubbles are unstable and
form  the time-dependent states \cite{CvSol,CG}.
\par
For charged bubbles there are extra positive terms:
contribution caused by the energy and mass of the external electromagnetic
field
\begin{equation}
E_{e.m.} = M_{e.m.} = \frac{e^2}{2r_0},
\label{EMem}
\end{equation}
and contribution to mass caused by gravitational field of the external
electromagnetic field ( determined by Tolman relation for
the external e.m. field)
\begin{equation}
M_{grav. e.m.} =  E_{e.m.} = \frac{e^2}{2r_0}.
\label{Mgrem}
\end{equation}
As a result the total energy for charged bubble is
\begin{equation}
E_{tot.bubble} = E_{wall} + E_{e.m.} = 4\pi r_0 ^2 \epsilon_{min}
+ \frac{e^2}{2r_0},
\label{Etot}
\end{equation}
and the total mass will be
\begin{equation}
M_{tot.bubble} = M_{0 bubble} + M_{e.m.} + M_{grav.e.m.} =
- E_{wall} + 2E_{e.m.} = -4\pi r_0 ^2 \epsilon_{min} + \frac{e^2}{r_0}.
\label{Mtot}
\end{equation}
Minimum of the total energy is achieved by
\begin{equation}
r_0=(\frac{e^2}{16\pi \epsilon_{min}})^{1/3},
\label{rmin}
\end{equation}
which yields the following expressions for total mass and energy of
the stationary state
\begin{equation}
M_{tot}^* =E_{tot}^* =\frac {3e^2}{4r_0}.
\label{EMst}
\end{equation}
One sees that the resulting total mass of charged bubble is positive,
however, due to negative contribution of $M_{0 bubble}$ it can be lower
than BPS energy bound of the domain wall forming this bubble.
This remarkable property of the bubble models (  `ultra-extreme'
states for the Type I domain walls in \cite {CvSol} ) allows one to overcome
BPS bound \cite{GH} and opens the way to get the ratio $m^2 \ll e^2$ which is
necessary for particle-like models.
\par
\subsection{Baglike source in supergravity}
\par
In supergravity the scalar potential has a more complicate form
\cite{WB,CvSol,CG,Mor2}
\begin{equation}
V_{sg}=e^{k^2 K}( K^{i\bar j}D_{i} W\overline {D_{j} W }
-3k^2 W \bar W),
\label{SGpot}
\end{equation}
where $K$ is K\"ahler potential $ K^{i\bar j}=\frac {\partial ^2 K}
{\partial \Phi _i \partial \bar \Phi _j}$, and
$k^2=8\pi G_N$, $G_N$ is the Newton constant.
In the small $kW$ limit, this expression
turns into potential of global susy. In this approximation, the
above treatment of the charged domain wall bubble will be valid in
supergravity.
The preserving supersymmetry vacuum state has
to satisfy the condition $D_i W \equiv W_i + k^2 K_i W =0$.
This condition is satisfied for the internal vacuum state II) only in the
limit $k^2 \to 0$ since  $W= \lambda m \eta^2/c$ inside
the bag, and $D_i W \approx k^2 K_i W $ there. In the order  $k^2$ the vacuum
state  II) does not preserve supersymmetry.
There appears also an extra contribution to stress-energy tensor
having the leading term
\begin{equation}
T_{\mu \nu} =3(k^2/8\pi) e^{k^2K}\vert W \vert ^2 g_{\mu\nu},
\label{Tsg}
\end{equation}
and yielding the negative cosmological constant $\Lambda =-3 k^4 e^{k^2K}
\vert W \vert ^2$ and to anti-de Sitter space-time for the bag interior.
General expression for cosmological constant inside the bag has the form
\begin{equation}
\Lambda=k^4 e^{k^2 K}\sum _{i}\{k^2 | K_i W |^2 - 3 |W|^2 \}.
\label{lamsg}
\end{equation}
It yields AdS vacuum if $k^2 | K_i W |^2 - 3 |W|^2 <0 $.
\par
In the same time the vacuum state I) in external region has $W=0$ and
$\Lambda =0$, and it preserves supersymmetry for strong chiral fields.
\par
\section{Conclusion}
A regularized source of the Kerr-Newman solution is considered having the
structure of a rotating bag with AdS interior and a smooth domain wall
boundary.
\par
We show that the Witten superconducting string model can be generalized and
adapted forming a charged supersymmetric superconducting bag with AdS
interior and with a long range external gauge field which is necessary for
description of charged black holes.
\par
Since 1989 a successive accumulation of evidences is observed relating
the structure of Kerr geometry with physics of elementary particles.

\section{Acknowledgement}
\par
We would like to thank J. Wess, H. Nicolai, S. Deser, M. Vasiliev,
Yu. Obukhov, G. Magli, G. Alekseev, S. Duplij, E. Elizalde,
S.R. Hildebrandt,
D. Bazeia and J. Morris for very useful discussions and comments.
We thank also organizers of the Advanced Research Workshop for kind
invitation and financial support.
\par

\end{document}